\documentclass[runningheads,a4paper]{llncs}

\usepackage{amssymb}
\usepackage{graphicx}
\usepackage{epstopdf}
\usepackage{bussproofs}
\usepackage{lscape}
\usepackage{pifont}
\usepackage{diagrams}
\usepackage{amsmath}
\usepackage{wasysym}
\usepackage{soul}
\usepackage{cmll}
\usepackage{wrapfig}
\usepackage{multicol}

\usepackage{url}
\urldef{\mailsa}\path|matthijs.vakar@ox.ac.uk|

\newcommand{\ra}[1]{\stackrel{#1}{\longrightarrow}}
\newcommand{\txt}[1]{\quad\textnormal{#1}\quad}
\newcommand{\sem}[2][M\!,g]{ [\![ #2 ]\!]^{}}

\newcommand{\B}[0]{\widetilde{\mathbb{B}_*}}
\newcommand{\N}[0]{\widetilde{\mathbb{N}_*}}
\newcommand{\ttt}[0]{\mathsf{tt}}
\newcommand{\fff}[0]{\mathsf{ff}}
\newcommand{\xmark}{\ding{55}}
\newcommand{\cmark}{\ding{51}}
\newcommand{\Gamecat}{\mathbf{Game}}
\newcommand{\DGame}{\mathbf{DGame}}
\newcommand{\str}{\mathsf{str}}
\newcommand{\wstr}{\mathsf{wstr}}
\renewcommand{\emph}[1]{\textbf{#1}}
\newcommand{\Osat}[0]{\textnormal{\textsf{O}-\textsf{sat}}}
\newcommand{\ob}[0]{\mathsf{ob}}
\newcommand{\tot}[0]{\smiley}

\begin{document}

\mainmatter  

\title{Games for Dependent Types}

\titlerunning{Games for Dependent Types}
%
%
\author{Samson Abramsky\inst{1}, Radha Jagadeesan\inst{2} and Matthijs V\'ak\'ar\inst{1}}
\authorrunning{Samson Abramsky, Radha Jagadeesan and Matthijs V\'ak\'ar}

\institute{University of Oxford, Oxford, UK\\
\and
DePaul University, Chicago, USA 
}

%
%

\toctitle{Games for Dependent Types}
\tocauthor{Samson Abramsky, Radha Jagadeesan and Matthijs V\'ak\'ar}
\maketitle

\vspace{-20pt}
\begin{abstract}
We present a model of dependent type theory (\textsf{DTT}) with $\Pi$-, $1$-, $\Sigma$- and intensional $\mathsf{Id}$-types, which is based on a slight variation of the category of AJM-games and history-free winning strategies. The model satisfies Streicher's criteria of intensionality and refutes function extensionality. The principle of uniqueness of identity proofs is satisfied.

We show it contains a submodel as a full subcategory which gives a faithful model of \textsf{DTT} with $\Pi$-, $1$-, $\Sigma$- and intensional $\mathsf{Id}$-types and, additionally, finite inductive type families. This smaller model is fully (and faithfully) complete with respect to the syntax   at the type hierarchy built without $\mathsf{Id}$-types, as well as at the class of types where we allow for one strictly positive occurrence of an $\mathsf{Id}$-type. Definability for the full type hierarchy with $\mathsf{Id}$-types remains to be investigated.
\end{abstract}
\vspace{-25pt}
\section{Introduction}
\vspace{-7pt}
Dependent Type theory (\textsf{DTT}) can be seen as the extension of the simple $\lambda$-calculus along the Curry-Howard correspondence from a proof calculus for (intuitionistic) propositional logic to one for predicate logic. It forms the basis of many proof assistants, like NuPRL, LEGO and Coq, and is increasingly being considered as a more expressive type system for programming, as implemented in e.g. ATS, Cayenne, Epigram, Agda and Idris. \cite{altenkirch2005dependent} A recent source of enthusiasm in this field is homotopy type theory (\textsf{HoTT}), which refers to an interpretation of \textsf{DTT} into abstract homotopy theory \cite{awodey2009homotopy} or, conversely, an extension of \textsf{DTT} that is sufficient to reproduce significant results of homotopy theory \cite{hottbook}. In practice, the latter means \textsf{DTT} with $\Sigma$-, $\Pi$-, $\mathsf{Id}$-types, a universe satisfying the \emph{univalence axiom}, and certain higher inductive types. The univalence axiom is an extensionality principle which implies, in particular, the axiom of function extensionality \cite{hottbook}.

Game semantics provides a unified framework for intensional, computational semantics of various type theories, ranging from pure logics~\cite{abramsky1994gamesll} to programming languages~\cite{hyland2000full,abramsky2000full,abramsky2005game} with a variety of features (e.g. non-local control \cite{laird1997full}, state \cite{abramsky1996linearity,abramsky1998fully,murawski2011game}, non-determinism \cite{harmer1999fully}, probability \cite{danos2002probabilistic}, dynamically generated local names \cite{abramsky2004nominal}) and evaluation strategies~\cite{abramsky1998cbvgames}. A game semantics for \textsf{DTT}  has,  surprisingly, so far been  absent.   
Our hope is that such a  semantics  will provide an alternative analysis of the implications of the subtle shades of intensionality that arise in the analysis of \textsf{DTT}~\cite{streicher1993investigations,hofmann1997syntax}. 
Moreover, the game semantics of \textsf{DTT} is based on very different, one might say orthogonal intuitions to those of the homotopical models: temporal  rather than spatial, and directly reflecting the structure of computational processes. One goal, to which we hope this work  will be a stepping stone, is a game semantics of \textsf{HoTT} doing justice to both the spatial and temporal aspects of identity types. Indeed, such an investigation might even lead to a computational interpretation of the univalence axiom which has long been missing, although a significant step in this direction was recently taken by the constructive cubical sets model of \textsf{HoTT}~\cite{bezem2014model}.  

Our game theoretic model of \textsf{DTT} is inspired in part by the domain model of \textsf{DTT} \cite{palmgren1990domain}.  This model views a type family as a continuous function to a domain of domains, a witness of a $\Pi$-type $\Pi_{x:A} B$ as a continuous (set theoretic) dependent function and interprets identity types via a kind of  intersection.    We follow this recipe for modelling type families  and identity types.   We adapt the viewpoint of  the game semantics of system F \cite{abramsky2005game}
to describe the $\Pi$-type to capture the intuitive idea that the specialisation of a term at type $\Pi_{x:A} B$ to a specific instance $B[a/x]$  is the responsibility solely of the context that provides the argument $a$ of type $A$; in contrast, any valid term of $\Pi_{x:A} B$ has to operate within the constraints enforced by the context.  Our definition draws its power from the fact that in a game semantics, these constraints are enforced not only on completed computations, but also on the incomplete computations that arise when a term interacts with its context.    Thus, while we follow the formal recipes of \cite{palmgren1990domain}, the temporal character of game semantics results in strikingly different properties of the resulting model.

In the rest of this paper, we describe a game theoretic model of \textsf{DTT} with $1$-, $\Sigma$-, $\Pi$- and intensional $\mathsf{Id}$-types, where (lists of dependent) AJM-games interpret types and (lists of) history-free winning strategies on games of dependent functions interpret terms. We next specialize to the semantic type hierarchy formed by the $1$-, $\Sigma$-, $\Pi$-constructions and substitution over a set of finite dependent games. We show that this gives of model of \textsf{DTT} which additionally supports finite inductive type families. Our two models have the following key properties.\vspace{-2pt}
\begin{itemize}
\item The place of the $\mathsf{Id}$-types in the intensionality spectrum compares as follows with the domain semantics and with HoTT.\vspace{8pt}\newline\resizebox{!}{35pt}{
\begin{tabular}{l||c|c|c}
 & \;\; Domains\;\; & \;\;HoTT\;\; &\;\; Games\;\;\\
 \hline
Failure of Equality Reflection  & \cmark & \cmark &  \cmark\\
Streicher Intensionality Criteria $(I1)$ and $(I2)$ \;& \cmark & \cmark & \cmark\\
Streicher Intensionality Criterion $(I3)$ & \xmark & \xmark & \cmark\\
Failure of Function Extensionality (\textsf{FunExt}) &\xmark & \xmark & \cmark\\
Failure of Uniqueness of Identity Proofs (\textsf{UIP})\;\; & \xmark & \cmark & \xmark
\end{tabular}}
\vspace{10pt}
\item We show that the smaller model faithfully models the terms of a version of \textsf{DTT} with $1$-, $\Sigma$-, $\Pi$- and $\mathsf{Id}$-types and finite inductive type families. Moreover, it is fully complete at the types which do not involve $\mathsf{Id}$ in their construction or which involve one strictly positive $\mathsf{Id}$-type as a subformula.   In contrast, the domain theoretic model of \cite{palmgren1990domain} is not (fully) complete or faithful. 
\end{itemize}

\vspace{-12pt}
\section{A Category of Games}\vspace{-7pt}
The idea behind game semantics is to model a computation by an alternating sequence of interactions (the play) between a program (Player) and its environment (Opponent), following some rules specified by its datatype (the game). In this translation, programs become Player strategies, while termination corresponds to a strategy being winning or beating all Opponents. The charm of this interpretation is that it not only fully captures the intensional aspects of a program but that it combines this with the structural clarity of a categorical model, thus interpolating between traditional operational and denotational semantics.

We assume the reader has some familiarity with the basics of categories of AJM-games and strategies, as described in \cite{abramsky2009game}, and only briefly recall the definitions. We define a category $\Gamecat$ which has as objects AJM-games.
\begin{definition}[Game] A \emph{game} $A$ is a tuple $(M_A,\lambda_A,P_A,\approx_A,W_A)$, where
\begin{itemize}\vspace{-7pt}
\item $M_A$ is a countable set of \emph{moves};
\item \mbox{\begin{diagram}M_A & \rTo^{\textnormal{\scriptsize$\lambda_A=\langle\lambda_A^{OP},\lambda_A^{QA}\rangle$}} & \{O,P\}\times \{Q,A\}\end{diagram}} is a function which indicates if a move is made by \emph{Opponent ($O$) or Player ($P$)} and if it is a \emph{Question} ($Q$) or an \emph{Answer} ($A$), for which we write $\overline{O}=P$, $\overline{P}=O$ and $M_A^O:={\lambda_A^{OP}}^{-1}(O)$, $M_A^P:={\lambda_A^{OP}}^{-1}(P)$, $M_A^Q:={\lambda_A^{QA}}^{-1}(Q)$ and $M_A^A:={\lambda_A^{QA}}^{-1}(A)$;
\item  $P_A\subseteq M_A^\circledast$ is a non-empty prefix-closed set of \emph{plays}, where $M_A^\circledast$ is the set of finite sequences of uniquely occurring moves, with the properties
\begin{itemize}
\item[(p1)] $s=at\Rightarrow a\in M_A^O$;
\item[(p2)] $\forall_i \lambda_A^{OP}(s_{i+1})=\overline{\lambda_A^{OP}(s_i)}$, where we write $s_i$ for the $i$-th move in $s$;
\item[(p3)] $\forall_{t\leq s}|t\upharpoonright_{M_A^A}|\leq |t\upharpoonright_{M_A^Q}|$.
\end{itemize}
Here, $\leq$ denotes the prefix order and $|s|$ the length of a sequence. Write $\mathsf{j}_{A,s}(m)$ for the last unanswered question preceding an answer $m$ in a play $s$, which we say $m$ answers. $\mathsf{j}_{A,s}$ will be used to enforce \emph{stack discipline}.
\item $\approx_A$ is an equivalence relation on $P_A$, satisfying
\begin{enumerate}
\item[(e1)] $s\approx_A t\Rightarrow \lambda^*_A(s)=\lambda^*_A(t)$;
\item[(e2)] $s\approx_A t\; \wedge \;s'\leq s\; \wedge \;t'\leq t \; \wedge \; |s'|=|t'|\; \Rightarrow s'\approx_A t'$;
\item[(e3)] $s\approx_A t\wedge sa\in P_A\Rightarrow \exists_bsa\approx_A tb$.
\end{enumerate}
Here, $\lambda_A^*$ is the extension of $\lambda_A$ to sequences. 
\item $W_A\subseteq P_A^\infty$ is a set of \emph{winning plays}, where $P_A^\infty$ is the set of infinite plays, i.e. infinite sequences of moves such that all their finite prefixes are in $P_A$, such that $W_A$ is closed under $\approx_A$ in the sense that\vspace{-5pt}
$$\left( s\in W_A\wedge t\notin W_A \right)\Rightarrow \exists_{s_0\leq s,t_0\leq t}|s_0|=|t_0|\wedge s_0\not\approx_A t_0. $$
\end{itemize}
\end{definition}
Our notion of morphism will be defined in terms of strategies on games.
\begin{definition}[Strategy] A \emph{strategy on $A$} is a subset $\sigma\subseteq P_A^{\mathsf{even}}$ satisfying\vspace{-3pt}
\begin{itemize}
\item[] (Causal Consistency): $sab\in \sigma\Rightarrow s\in \sigma$;
\item[] (Representation Independence): $ s\in\sigma\;\wedge\; s\approx_A t\Rightarrow t\in\sigma$;
\item[] (Determinacy): $sab,ta'b'\in \sigma\;\wedge\; sa\approx_A ta'\Rightarrow sab\approx_A ta'b'$.
\end{itemize}
\end{definition}
We write $\str(A)$ for the set of strategies on $A$. We sometimes identify $\sigma$ with the subset of $P_A$ that is obtained as its prefix closure. In fact, we restrict to history-free strategies, as we are modelling computation without mutable state. 
\begin{definition}[History-Free Strategy] We call a strategy $\sigma\in\str(A)$ \emph{history-free}, if there exists a non-empty causally consistent subset $\phi\subseteq \sigma$ (called a \emph{history-free skeleton}) such that
\begin{itemize}\vspace{-6pt}
\item[] (Uniformization): $\forall_{sab\in\sigma}s\in\phi\Rightarrow\exists !_{b'}sab'\in\phi$;
\item[] (History-Freeness 1): $sab,tac\in\phi\Rightarrow b=c$;
\item[] (History-Freeness 2): $\left(sab,t\in\phi\;\wedge\;ta\in P_A\right)\Rightarrow tab\in\phi$.\vspace{-4pt}
\end{itemize}
Then, $\phi$ is induced by a partial function on moves and $\sigma=\{t \;|\; \exists_{s\in\phi}t\approx_A s\}$.
\end{definition}
From now on, we assume strategies to be history-free. Winning conditions give rise to the notion of a winning strategy, the semantic equivalent of a normalising or total term. A winning strategy always has a response to any valid $O$-move. Furthermore, if the result of the interaction between a strategy and Opponent is an infinite play, then this is a member of the set of winning plays.
\begin{definition}[Winning Strategy]A strategy $\sigma\in\str(A)$ is \emph{winning} if it satisfies
\begin{itemize}
\item[] (Finite Wins): If $s$ is $\leq$-maximal in $\sigma$, then $s$ is $\leq$-maximal in $P_A$.
\item[] (Infinite Wins): If $s_0\leq s_1\leq \ldots$ is an infinite chain in $\sigma$, then $\bigcup_i s_i\in W_A$.
\end{itemize}
\end{definition}
We write $\wstr(A)$ for the set of winning strategies on $A$.
Next, we define some constructions on games, starting with their symmetric monoidal closed structure.
\begin{definition}[Tensor Unit] We define the game $I:=(\emptyset,\emptyset,\{\epsilon\},\{(\epsilon,\epsilon)\},\emptyset)$.\end{definition}

\begin{definition}[Tensor] For games $A,B$, we define\newline $A\otimes B:=(M_A+M_B=\Sigma_{i\in \{A,B\}}M_i,[\lambda_A,\lambda_B],P_{A\otimes B},\approx_{A\otimes B},W_{A\otimes B})$ with
\begin{itemize}
\item \resizebox{\linewidth}{!}{$P_{A\otimes B}=\{s\;|\; s\upharpoonright_A \in P_A \wedge s\upharpoonright_B \in P_B \wedge\linebreak \mathsf{fst}^*(\mathsf{j}_{A\otimes B,s}^*(s\upharpoonright_{M_{A\otimes B}^A}))=\mathsf{fst}^*(s\upharpoonright_{M_{A\otimes B}^A}) \}$;}
\item $s\approx_{A\otimes B} t:= s\upharpoonright_A \approx_A t\upharpoonright_A \;\wedge \; s\upharpoonright_B\approx_B t\upharpoonright_B\;\wedge\; \forall_{1\leq i\leq |s|} s_i\in M_A\Leftrightarrow t_i\in M_A$;
\item $W_{A\otimes B} :=\{s\in P_{A\otimes B}^\infty|\left(s\upharpoonright_A\in P_A^\infty \hspace{-2pt}\Rightarrow s\upharpoonright_A\in W_A\right)\wedge\left( s\upharpoonright_B\in P_B^\infty\hspace{-1pt} \Rightarrow s\upharpoonright_B\in W_B\right)\}$.
\end{itemize}
\end{definition}
\begin{definition}[Linear Implication] For games $A,B$, we define\\ $A\multimap B:=(M_A+M_B=\Sigma_{i\in \{A,B\}}M_i,[\overline{\lambda_A},\lambda_B],P_{A\multimap B},\approx_{A\multimap B},W_{A\multimap B})$ with
\begin{itemize}
\item \resizebox{\linewidth}{!}{$P_{A\multimap B}=\{s\;|\; s\upharpoonright_A \in P_A \wedge s\upharpoonright_B \in P_B \wedge\mathsf{fst}^*(\mathsf{j}_{A\multimap B,s}^*(s\upharpoonright_{M_{A\multimap B}^A}))=\mathsf{fst}^*(s\upharpoonright_{M_{A\multimap B}^A}) \}$;}
\item $s\approx_{A\multimap B} t:= s\upharpoonright_A \approx_A t\upharpoonright_A \;\wedge \; s\upharpoonright_B\approx_B t\upharpoonright_B\;\wedge\; \forall_{1\leq i\leq |s|} s_i\in M_A\Leftrightarrow t_i\in M_A$;
\item $W_{A\multimap B}:=\{s\in P_{A\multimap B}^\infty\;|\; s\upharpoonright_A\in W_A \Rightarrow s\upharpoonright_B\in W_B\}$.
\end{itemize}
\end{definition}
Note that the definitions of $\lambda_{-}$ imply that in $A\otimes B$ only Opponent can switch between $A$ and $B$, while in $A\multimap B$ only Player can. These definitions on objects extend to strategies, e.g. for (winning) strategies $\sigma\in\str(A), \tau\in\str(B)$, we can define a (winning) strategy $\sigma\otimes \tau=\{s\in P_{A\otimes B}^\mathsf{even}\;|\; s\upharpoonright_A\in \sigma \;\wedge\; s\upharpoonright_B\in \tau\}\in\str(A\otimes B)$. This gives us a model of multiplicative intuitionistic linear logic, with all structural morphisms consisting of appropriate variants of copycat strategies.
\begin{theorem}[Linear Category of Games] We define a category $\Gamecat$ by\vspace{-6pt}
\begin{itemize}
\item $\mathsf{ob}(\Gamecat):=\{A\;|\; A\;\textnormal{ is an AJM-game}\}$;
\item $\Gamecat(A,B):=\wstr(A\multimap B)$;
\item $\mathsf{id}_A:=\{s\in P_{A\multimap A}\;|\; s\upharpoonright_{A^{(1)}}\approx_A s\upharpoonright_{A^{(2)}}\}$, the \emph{copycat strategy} on $A$;
\item for $A\ra{\sigma}B\ra{\tau}C$, the composition (or \emph{interaction}) $A\ra{\sigma;\tau}C$ is defined from parallel composition $\sigma||\tau:=\{s\in M_{(A\multimap B)\multimap C}^\circledast \;|\; s\upharpoonright_{A,B}\;\in\sigma\;\wedge \; s\upharpoonright_{B,C}\;\in\tau\}$ plus hiding: $\sigma;\tau:=\{s\upharpoonright_{A,C}\;|\; s\in \sigma||\tau\}$.
\end{itemize}
Then, $(\Gamecat,I,\otimes,\multimap)$ is, in fact, a symmetric monoidal closed category.
\end{theorem}
To make this into a model of intuitionistic logic, a Cartesian closed category (ccc), through the (first) Girard translation, we need two more constructions on games, to interpret the additive conjunction $\&$ and exponential $!$, respectively. A play in $!A$ consists of any number of interleaved threads of plays in $A$.
\begin{definition}[With] We define the game\\ $A\& B:=(M_A+M_B,[\lambda_A,\lambda_B],P_A+P_B,\approx_A+\approx_B,W_A + W_B)$.
\end{definition}
\begin{definition}[Bang] We define $!A:=(\mathbb{N}\times M_A,\lambda_A\circ\mathsf{snd}, P_{!A},\approx_{!A},W_{!A})$ with\vspace{-3pt}
\begin{itemize}
\item $P_{!A}=\{s\;|\; \forall_{i\in\mathbb{N}}s\upharpoonright_i\in P_A\;\wedge\;  \mathsf{fst}^*(\mathsf{j}_{!A,s}^*(s\upharpoonright_{M_{!A
 }^A}))=\mathsf{fst}^*(s\upharpoonright_{M_{!A}^A}~)\}$; 
\item $s\approx_{!A} t:=\exists_{\pi\in S(\mathbb{N})}\forall_{i\in \mathbb{N}}s\upharpoonright_i\approx_A t\upharpoonright_{\pi(i)}\; \wedge \; (\pi\circ \mathsf{fst})^*(s)=\mathsf{fst}^*(t)$;
\item $W_{!A}:= \{s\in P_{!A}^\infty\;|\; \forall_i s\upharpoonright_i\in P_A^\infty \Rightarrow s\upharpoonright_i\in W_A\}$.
\end{itemize}
\end{definition}
Next, we note that $!$ can be made into a co-monad by defining, for $A\ra{\sigma}B$, \vspace{-3pt}
$$!\sigma:=\{s\in P_{!A\multimap !B}^\mathsf{even}\;|\; \exists_{\pi\in S(\mathbb{N})} \forall_{i\in \mathbb{N}} s\upharpoonright_{(\pi(i),A), (i,B)}\in \sigma\},\vspace{-3pt}$$
and natural transformations $$!A\ra{\mathsf{der}_A}A:=\{s\in P_{!A\multimap A}^\mathsf{even}\;|\; \exists_{i\in\mathbb{N}} s\upharpoonright_{!A}\upharpoonright_i\approx_A s\upharpoonright_A\}\txt{and}\vspace{-3pt}$$\vspace{-8pt} $$!A\ra{\delta_A}!!A:=\{s\in P_{!A\multimap !!A}^\mathsf{even}\;|\; \exists_{p:\mathbb{N}\times\mathbb{N}\hookrightarrow\mathbb{N}} \forall_{i,j\in \mathbb{N}} s\upharpoonright_{!A}\upharpoonright_{p(i,j)}\approx_A s\upharpoonright_{!!A}\upharpoonright_i\upharpoonright_j \}.$$

This allows us to define the co-Kleisli category $\Gamecat_!$, which has the same objects as $\Gamecat$, while $\Gamecat_!(A,B):=\Gamecat(!A,B)$. We have a composition $(f,g)\mapsto f^\dagger;g$, where we write $f^\dagger:=\delta_{\mathsf{dom}(f)};!(f)$, for which the strategies $\mathsf{der}_A$ serve as identities. We can define finite products in $\Gamecat_!$ by $I$ and $\&$ and write\vspace{-5pt}\newline \resizebox{\linewidth}{!}{\parbox{1.1\linewidth}{$$\mathsf{diag}_A:=\{s\in P_{!A\multimap (A\& A)}^\mathsf{even}\;|\; \exists_{i\in\mathbb{N}}(s=\epsilon)\vee(s\upharpoonright_{!A}\upharpoonright_i\approx_A s\upharpoonright_{A^{(1)}}\neq \epsilon)\vee (s\upharpoonright_{!A}\upharpoonright_i\approx_A s\upharpoonright_{A^{(2)}}\neq \epsilon)\}\vspace{-5pt}$$}} for the diagonal $!A\ra{} A\& A$. Moreover, we have Seely-isomophisms $!I\cong I$ and $!(A\&B)\cong !A\otimes !B,$ so we obtain a linear-non-linear adjunction $\Gamecat\leftrightarrows\Gamecat_!$, hence a model of multiplicative exponential intuitionistic linear logic. In particular, by defining $A\Rightarrow B:=!A\multimap B$, we get a ccc. We write $\mathsf{comp}_{A,B,C}$ for the internal composition $((A\Rightarrow B) \;\&\; (B\Rightarrow 
C))\ra{} A\Rightarrow C$ in $\Gamecat_!$.
\begin{theorem}[Intuitionist Category of Games]$(\Gamecat_!,I,\&,\Rightarrow)$ is a ccc.\end{theorem}
Note that for the hierarchy of intuitionistic types $A$ that are formed by operations $I$, $\&$ and $\Rightarrow$ from finite games,  winning strategies are the total strategies - strategies which respond to any $O$-move - for which infinite chattering can only occur because Opponent opens infinitely many threads of the same game.
 \vspace{-5pt}

\section{Dependent Games}\label{sec:depgam}\vspace{-7pt}
The previous section sketched how $\Gamecat_!$ models simple intuitionistic type theory. Next, we show how it comes equipped with a notion of dependent type. This leads to an indexed ccc $\DGame_!$ of dependent games and strategies.

We define a poset $\Gamecat_\trianglelefteq$ of games with $A\trianglelefteq B:=(M_A= M_B)\;\wedge\;(\lambda_B|_{M_A}=\lambda_A)\;\wedge\;(P_A\subseteq P_B)\;\wedge\; (s\approx_A t\quad \Leftrightarrow \quad s\in P_A\;\wedge \; s\approx_B t)\;\wedge\;(W_A=W_B\cap P_A^\infty)$. Given a game $C$, we define the cpo $\mathsf{Sub}(C)$ as the poset of its $\trianglelefteq$-subgames. We note that, for $A,B\in\mathsf{Sub}(C)$, $A\trianglelefteq B \Leftrightarrow P_A\subseteq P_B$.

For a game $A$, we define the set $\mathsf{ob}(\DGame_!(A))$ of \emph{games with dependency on $A$} as the set of continuous functions $\str(A)\ra{B}\mathsf{Sub}(\smiley B)$ for some other game $\smiley B$.  We note that $\mathsf{ob}(\DGame_!(I))$ is the set of pairs $A=(A(\bot),\smiley { {A}})$ where $A(\bot)\trianglelefteq \smiley {A}$, of which $\mathsf{ob}(\Gamecat_!)$ arises as the proper subset of diagonal elements $(A,A)$. We define more generally $\mathsf{ob}(\DGame_!(A)):=\mathsf{ob}(\DGame_!(\smiley A))$. Writing $s\mapsto \overline{s}$ for the function $P_{!A}\ra{}\mathcal{P}(P_A)$ inductively defined on the empty play, Opponent moves and Player moves, respectively, as  $\epsilon\mapsto \emptyset,\;\; s(i,a)\mapsto~\overline{s}, \quad s(i,a)(i,b)\mapsto \overline{s(i,a)}\cup \{t\;|\; \exists_{s'\in\overline{s}}t\approx_A s'ab \}$, we define the dependent function space as follows.
\begin{definition}[$\Pi$-Game]
Given $B\in\mathsf{ob}(\DGame_!(A))$, we define the $\Pi$-game $(\Pi_{A}B)(\bot)\;\trianglelefteq \; \smiley A\Rightarrow \smiley B$  inductively as \\
\resizebox{\linewidth}{!}{\parbox{\linewidth}{\vspace{-10pt}\begin{align*}
&\{\epsilon\}\;\bigcup
\{sa\;|\; s\in P_{(\Pi_{A} B)(\bot)}^\mathsf{even}\;\wedge\;  \exists_{\overline{sa\upharpoonright_{!\smiley { A}}}\subseteq \tau\in \wstr(A(\bot))} sa\in P_{A(\bot)\Rightarrow B(\tau)}\;\}\;\bigcup\\
&\{sab\;|\; sa \in P_{(\Pi_{A} B)(\bot)}^\mathsf{odd}\;\wedge\; \forall_{ \overline{sab\upharpoonright_{!\smiley { A}}}\subseteq \tau\in \wstr(A(\bot))}  sa\in P_{A(\bot)\Rightarrow B(\tau)}\Rightarrow sab\in P_{A(\bot)\Rightarrow B(\tau)}\;\}.
\end{align*}
}}
\vspace{-5pt}
\end{definition}\vspace{-12pt}

We note that we can make $\DGame_!(A )$ into a ccc by defining $I$ and $\&$ pointwise on dependent games $B$, while also performing the operation on $\smiley { B}$, and by defining $\smiley { B\Rightarrow C}:=\smiley { B}\Rightarrow \smiley {C}$ and $P_{(B\Rightarrow C)(\sigma)}:=\{s\in P_{B(\sigma)\Rightarrow C(\sigma)}\;|\; \exists_{\tau\wstr(B(\sigma))}\linebreak \overline{s\upharpoonright_{B(\sigma)}}\subseteq\tau\;\}$. This lets us define $\DGame_!( A)( B,C):= \wstr(\Osat(\Pi_{A}(B\Rightarrow C)))$ with the obvious identities and composition, which we discuss later. Here, the game $\Osat(A(\bot),\smiley { A})$ has plays $\{\epsilon\}\;\bigcup \{sa \in P_{\smiley { A}}^{\mathsf{odd}}\;|\; s\in P^{\mathsf{even}}_{\Osat(A(\bot),\smiley { A})}\;\}\;\bigcup\;$\linebreak$
\{sab\in P^{\mathsf{even}}_{\smiley { A}}\;|\; sa\in P^{\mathsf{odd}}_{\Osat(A(\bot),\smiley { A})}\;\wedge\; (sa\in P_{A(\bot)}\Rightarrow sab\in P_{A(\bot)})\}$. Explicitly, we have the \emph{game $\Osat(\Pi_{A} B)$ of dependent functions from $A$ to $B$}\\
\resizebox{\linewidth}{!}{\parbox{\linewidth}{\vspace{-10pt}
\begin{align*}
&\{\epsilon\}\;\bigcup \{sa\;|\; s\in P_{\Osat(\Pi_{A} B)}^\mathsf{even} \;\}\;\bigcup\\
&\{sab\;|\; sa \in P_{\Osat(\Pi_{A} B)}^\mathsf{odd}\;\wedge \;\forall_{ \overline{sab\upharpoonright_{!\smiley { A}}}\subseteq \tau\in \wstr(A(\bot))} sa\in P_{A(\bot)\Rightarrow B(\tau)}\Rightarrow sab\in P_{A(\bot)\Rightarrow B(\tau)}\;\}.
\end{align*}}}
\vspace{-10pt}

Following the mantra of game semantics for quantifiers \cite{abramsky2005game}, in $\Osat(\Pi_{A} B)$, Opponent can choose a winning strategy $\tau$ on $A(\bot)$ while Player has to play in a way that is compatible with all choices of $\tau$ that have not yet been excluded. Similarly to the approach taken in the game semantics for polymorphism \cite{abramsky2005game}, we do not specify all of $\tau$ in one go, as this would violate ``Scott's axiom'' of continuity of computation. Instead, $\tau$ is gradually revealed, explicitly so by playing in $!\smiley { A}$ and implicitly by playing in $\smiley { B}$. That is, unless Opponent behaves naughtily, in the sense that there is no winning history-free strategy $\tau$ on $A(\bot)$ which is consistent with her behaviour while $s\upharpoonright_{\smiley { B}}$ obeys the rules of $B(\tau)$. In case of such a naughty Opponent, any further play in $\smiley { A}\Rightarrow \smiley { B}$ is permitted.

For an example, let $\mathsf{days}(n):=\{m\;|\; \textnormal{there are $>m$ days in the year $n$}\}$ and define $\widetilde{\mathsf{days}_*}(\bot)=\widetilde{\emptyset_*}$, $\widetilde{\mathsf{days}_*}(n):=\widetilde{\mathsf{days}(n)_*}$ to obtain a game depending on $\N$ (with $\widetilde{\mathsf{days}}(n)=\widetilde{\mathbb{N}_{<365}{}_*}\textnormal{ or }\widetilde{\mathbb{N}_{<366}{}_*}$). Here, $\widetilde{X_*}$ signifies the  game with $P_{\widetilde{X_*}}=\{\epsilon,*\}\cup\{ *x\;| x\in X\}$ and $\approx_X=\mathsf{id}_X$  Then, the following are valid strategies.
\begin{figure}
\centering\resizebox{\linewidth}{!}{
$\begin{array}{c|c|c|c||c}
\begin{array}{ccc}
!\N &\quad & \widetilde{\mathsf{days}_*}\\
\hline
 & & *\\
 & & 364\\
 & & \\
 & & \\
 & & \\
 & & \\
\end{array}
\hspace{10pt}&\hspace{10pt}
\begin{array}{ccc}
!\N &\quad & \widetilde{\mathsf{days}_*}\\
\hline
 & & *\\
(i,*) & & \\
(i,1984) & & \\
 & & 365 \\
 & & \\
 & & \\
\end{array}
\hspace{10pt}& \hspace{10pt}
\begin{array}{ccc}
!\N &\quad & \widetilde{\mathsf{days}_*}\\
\hline
 & & *\\
(i,*) & & \\
(i,1984) & & \\
(i+1,*) & & \\
(i+1,1985) & & \\
 & & 365 \\
\end{array}
\hspace{10pt}& \hspace{10pt}
\begin{array}{ccccc}
!\N &\quad & !\widetilde{\mathsf{days}_*} & \quad & \widetilde{\mathsf{days}_*}\\
\hline
 & & & & *\\
 & &(i,*) &  & \\
 & &(i,m) &  & \\
 & & & & m\\
 & & & &\\
 & & & &
\end{array}
 \hspace{10pt}&
\begin{array}{c}
\vspace{4pt}\\
O\\
P\\
O\\
P\\
O\\
P
\end{array}
\end{array}
$}\vspace{-5pt}
\caption{Three strategies on $\Osat(\Pi_{!\N}\widetilde{\mathsf{days}_*})$ and one on $\Osat(\Pi_{!\N}!\widetilde{\mathsf{days}_*}\multimap \widetilde{\mathsf{days}_*})$. The first as all years have $> 364$ days, the second as $1984$ was a leap year, the third as Player can play any move in $\smiley\widetilde{\mathsf{days}}_* =\widetilde{\mathbb{N}_{<366}{}_*}$ after Opponent has not played along a strategy on $\widetilde{\mathbb{N}}_*$ and the fourth as Opponent makes the move $m$ first, after which Player can safely copy it. In the paired moves, Player chooses an (irrelevant)  index $i$.}\vspace{-17pt}
\end{figure}\;\\
 The fourth example is especially important, as it generalises to a (derelicted) $B$-copycat on $\Osat(\Pi_{!A}(!B\multimap B))$ for arbitrary $B$, denoted $\mathbf{v}_{[A],[B]}$ in section \ref{sec:ctxt}. This~motivates why Opponent can narrow down the fibre of $B$ freely, while Player cannot. To see that Player should not be able to narrow down the fibre of $B$, note that we do not want $f:=\{\epsilon,*365\}$ to define a strategy on $\Osat(\Pi_{!\N}\widetilde{\mathsf{days}_*})$, as $1983;f=\{\epsilon,*365\}\notin\str(\widetilde{\mathsf{days}_*(1983)})$.
\begin{theorem}We obtain a strict indexed ccc \mbox{\small
\begin{diagram}
\DGame_!(I)^{op} & \rTo^{(\DGame_!,-\{-\})} & \mathsf{Cat}
\end{diagram}}
\linebreak of dependent games, if we define
\begin{itemize}
\item fibrewise object sets $\ob(\DGame_!(A)):=\{\str(\smiley { A})\ra{B}\mathsf{Sub}(\smiley { B})\;|\; \smiley { B}\in \ob(\Gamecat_!)\;\wedge\; B\;\textnormal{continuous}\;\}$;
\item fibrewise hom-sets $\DGame_!(A)(B,C):=\wstr(\Osat(\Pi_{!A}(!B\multimap C)))$;
\item fibrewise identities $\mathsf{der}_B:= \{s\in P_{\Osat(\Pi_{!A}(!B\multimap B))}\;|\; \exists_i  s\upharpoonright_{!B}\upharpoonright_i\approx_B s\upharpoonright_{B}\}$;
\item if $B\ra{\tau}C\ra{\tau'}D\in\DGame_!(A)$, $\tau^\dagger;_A \tau':=\mathsf{diag}^\dagger_{A};\tau^\dagger \otimes \tau' ; \mathsf{comp}_{\smiley B,\smiley C,\smiley D}$;
\item  given $f\in \Gamecat_!(A',A)$, we define the change of base functor $-\{f\}$: $B\{f\}\in \mathsf{ob}(\DGame_!(A'))$ where $B\{f\}(\sigma):=B(!(\sigma);f)$ and $\smiley { B\{f\}}:=\smiley { B}$ and $\tau\{f\}:=f^\dagger ; \tau$.
\end{itemize}

\end{theorem}

Seeing that $\DGame_!(I)$ additionally has a terminal object $I$ to interpret the empty context, we are well on our way to producing a  model of dependent type theory \cite{vakar2015syntax}: we only need to interpret context extension. This takes the form of the comprehension axiom for $\DGame_!$, which states that for each $A\in\mathsf{ob}(\DGame_!(I))$ and $B\in\mathsf{ob}(\DGame_!(A))$ the following presheaf is representable\vspace{-2pt} $$x\mapsto\DGame_!(\mathsf{dom}(x))(I,B\{x\}):(\DGame_!(I)/A)^{op}\ra{}\mathsf{Set}.$$ 
Unfortunately, this fails, as $\DGame_!(I)$ does not yield a sound interpretation of dependent contexts. Essentially, the problem is that we do not have \emph{additive $\Sigma$-types}, appropriate generalisations $\Sigma_A^\& B$ of $\&$ to interpret dependent context extension in $\DGame_!(I)$.
\begin{theorem}$\DGame_!$ does not satisfy the comprehension axiom.
\end{theorem}

\vspace{-8pt}
\section{A Category with Families of Context Games}\vspace{-7pt} \label{sec:ctxt}
All is not lost, however. In fact, we have almost translated the structural core of the syntax of \textsf{DTT} into the world of games and strategies. The remaining generalisation, necessitated by the lack of additive $\Sigma$-types, is to dependent games depending on multiple (mutually dependent) games. We can produce a categorical model of \textsf{DTT} out of the resulting structure by applying a so-called \emph{category of contexts ($\mathsf{Ctxt}$) construction}, which is precisely how one builds a categorical model from the syntax of dependent type theory \cite{hofmann1997syntax,pitts1995categorical}. This can be seen as a way of making our indexed category satisfy the comprehension axiom, extending its base category by (inductively) adjoining (strong) $\Sigma$-types formally, analogous to the $\mathsf{Fam}$-construction of \cite{abramsky1998cbvgames} which adds formal co-products.

The problem which needs to be addressed is how to interpret dependent types and dependent functions of more variables. This is done through a notion of context game and a generalisation of the $\Pi$-game construction from section~\ref{sec:depgam}.
\vspace{-10pt}
\begin{definition}[Context Game]
We define a \emph{context game} $[X_i]_{1\leq i \leq n}$ to be a list  where $X_{i}$ is a game with dependency on $[X_j]_{j< i}$, i.e. a continuous function $\str(\tot{ X_1})\times\cdots\times\str(\tot{ X_{i-1}})\ra{X_i}\mathsf{Sub}(\tot{ X_i})$ for some game $\tot{ X_i}$.
\end{definition}
\vspace{-5pt}
\begin{definition}[Dependent $\Pi$-game]For a game $X_{n+1}$ depending on $[X_i]_{i\leq n}$, we define the game $\Pi_{X_n}X_{n+1}$ depending on $[X_i]_{i\leq n-1}$ by $\tot{ \Pi_{X_n}X_{n+1}}:=\tot{ X_n}\Rightarrow \tot{ X_{n+1}}$ from which $(\Pi_{X_n}X_{n+1})(\sigma_1,\ldots,\sigma_{n-1})$ is carved out as\newline\resizebox{\linewidth}{!}{ \parbox{\linewidth}{\vspace{-10pt}
\begin{align*}
&\{\epsilon\}\;\bigcup \{sa\;|\; s\in P_{(\Pi_{X_n}X_{n+1})(\sigma_1,\ldots,\sigma_{n-1})}^\mathsf{even} \;\wedge\; \exists_{\overline{sa\upharpoonright_{!\tot{ X_n}}}\subseteq\tau\in\wstr(X_n(\sigma_1,\ldots,\sigma_{n-1}))}\\
&sa\in P_{X_n(\sigma_1,\ldots,\sigma_{n-1})\Rightarrow X_{n+1}(\sigma_1,\ldots,\sigma_{n-1},\tau)}\;\}\;\bigcup \{sab\;|\; sa \in P_{(\Pi_{X_n}X_{n+1})(\sigma_1,\ldots,\sigma_{n-1})}^\mathsf{odd}\\
&\wedge \forall_{\overline{sab\upharpoonright_{!\tot{ X_n}}}\subseteq\tau\in\wstr(X_n(\sigma_1,\ldots,\sigma_{n-1}))}sa\in P_{X_n(\sigma_1,\ldots,\sigma_{n-1})\Rightarrow X_{n+1}(\sigma_1,\ldots,\sigma_{n-1},\tau)}\Rightarrow\\& sab\in P_{X_n(\sigma_1,\ldots,\sigma_{n-1})\Rightarrow X_{n+1}(\sigma_1,\ldots,\sigma_{n-1},\tau)}\;\}.
\end{align*}}}\vspace{-10pt}
\end{definition}
Consequently, the \emph{game of dependent functions of multiple arguments} $\Osat(\Pi_{X_1}\cdots\Pi_{X_n}X_{n+1})$ is carved out in $\tot{ X_1}\Rightarrow\cdots\Rightarrow\tot{ X_n}\Rightarrow \tot{ X_{n+1}}$ as\newline
\resizebox{\linewidth}{!}{ \parbox{\linewidth}{\vspace{-5pt}
\begin{align*}
&\{\epsilon\}\;\bigcup\;\{sa\;|\; s\in P_{\Osat(\Pi_{X_1}\cdots\Pi_{X_n}X_{n+1})}^\mathsf{even} \;\}\;\bigcup \; \{sab\;|\; sa \in P_{\Osat(\Pi_{X_1}\cdots\Pi_{X_n}X_{n+1})}^\mathsf{odd}\wedge\\
& \forall_{\overline{sab\upharpoonright_{!\tot{ X_1}}}\subseteq \tau_1\in\wstr(X_1(\bot))}\cdots\forall_{\overline{sab\upharpoonright_{!\tot{ X_n}}}\subseteq \tau_n\in \wstr(X_n(\tau_1,\ldots,\tau_{n-1}))}
\\
&  sa\in P_{X_1(\bot)\Rightarrow\cdots\Rightarrow X_{n+1}(\tau_1,\ldots,\tau_n)}\Rightarrow sab\in P_{X_1(\bot)\Rightarrow\cdots \Rightarrow X_{n+1}(\tau_1,\ldots,\tau_n)}\;\}.
\end{align*}}}\vspace{-5pt}

For illustration, define a game $\widetilde{\mathsf{RA}_*}$ depending on the context game $[\N,\widetilde{\mathsf{days}_*}]$ by $\mathsf{RA}(n,m):=\{\textnormal{Rick Astley lyrics from songs released before day $m$ of year $n$}\}$.\vspace{-10pt}
\begin{figure}[htb]\vspace{-10pt}
\centering\resizebox{\linewidth}{!}{
$
\begin{array}{c|c||c}
\begin{array}{ccc}
!\N & \widetilde{!\mathsf{days}_*} &\widetilde{\mathsf{RA}_*} 
\\
\hline
&&*\\
&(i,*)&\\
&(i,m>206) & \\
(j,*) && \\
(j,1987) &&\\
 & & \textnormal{Never Gonna Give You Up}
\end{array}
&
\begin{array}{ccc}
!\N & \widetilde{!\mathsf{days}_*}& \widetilde{\mathsf{RA}_*} 
\\
\hline
&&*\\
(i,*) && \\
(i,n>1987) &&\\
 & & \textnormal{Never Gonna Let You Down}\\
 & & \\
 &&
\end{array}
&
\begin{array}{c}
\vspace{5pt}\\
O\\
P\\
O\\
P\\
O\\
P
\end{array}
\end{array}
$
}
\caption{Two examples of (partial) strategies on $\Osat(\Pi_{!\N}\Pi_{!\widetilde{\mathsf{days}_*}}\widetilde{\mathsf{RA}_*})$.}\vspace{-30pt}
\end{figure}

We define a category $\mathsf{Ctxt}(\DGame_!)$ with objects context games and morphisms which are defined inductively as (dependent) lists of winning strategies on appropriate games of dependent functions. We show that this has the structure of a category with families (CwF) \cite{hofmann1997syntax}, a canonical notion of model of \textsf{DTT}. This gives a more concise presentation of the resulting indexed category with comprehension, where we also add formal $\Sigma$-types in the fibres.

\begin{definition}[CwF] A CwF is a category $\mathcal{C}$ with a terminal object $\cdot$, for all objects $\Gamma$ a set $\mathsf{Ty}(\Gamma)$, for all $A\in \mathsf{Ty}(\Gamma)$ a set $\mathsf{Tm}(\Gamma,A)$, for all $\Gamma'\ra{f}~\Gamma$ in $\mathcal{C}$ functions $\mathsf{Ty}(\Gamma)\ra{-\{f\}}\mathsf{Ty}(\Gamma')$ and $\mathsf{Tm}(\Gamma,A)\ra{-\{f\}}\mathsf{Tm}(\Gamma',A\{f\})$, such that
\begin{tabular}{llll}
$A\{\mathsf{id}_\Gamma\}=A$ & (Ty-Id)\hspace{60pt} & $A\{g\circ f\}=A\{g\}\{f\}$\hspace{23pt} & (Ty-Comp)\\
$t\{\mathsf{id}_\Gamma\}=A$ \hspace{23pt} & (Tm-Id)& $t\{g\circ f\}=t\{g\}\{f\}$ & (Tm-Comp),
\end{tabular}\\
for $A\in\mathsf{Ty}(\Gamma)$ a morphism $\Gamma.A\ra{\mathbf{p}_{\Gamma,A}}\Gamma$ of $\mathcal{C}$ and $\mathbf{v}_{\Gamma,A}\in\mathsf{Tm}(\Gamma.A,A\{\mathbf{p}_{\Gamma,A}\})$ and, finally, for all $t\in \mathsf{Tm}(\Gamma',A\{f\})$ a morphism $\Gamma'\ra{\langle f,t\rangle}\Gamma.A$ such that\linebreak
\begin{tabular}{llll}
$\mathbf{p}_{\Gamma,A}\circ \langle f,t\rangle=f$ & (Cons-L)\hspace{35pt} & 
$\mathbf{v}_{\Gamma,A}\{\langle f,t\rangle\}=t $&(Cons-R)  \\
$\langle \mathbf{p}_{\Gamma,A},\mathbf{v}_{\Gamma,A}\rangle=\mathsf{id}_{\Gamma.A}$ \hspace{7pt} &(Cons-Id) &
$\langle f,t\rangle \circ g=\langle f\circ g,t\{g\}\rangle$ \hspace{7pt} & (Cons-Nat).
\end{tabular}

\end{definition}

\begin{theorem}We have a CwF $(\mathsf{Ctxt}(\DGame_{!}),\mathsf{Ty},\mathsf{Tm},\mathbf{p},\mathbf{v}, -.-, \langle - ,-\rangle)$.
\end{theorem}We define the required structures. All equations follow trivially from the definitions and the two claims stated.
We define $\mathsf{Ty}([X_i]_i)$ as the set of \emph{context games with dependency} on $[X_i]_i$: $[Y_j]_j\in\mathsf{Ty}([X_i]_i)$ iff $[X_i]_i.[Y_j]_j:=[X_1,\ldots,X_n,Y_1,\ldots,Y_m]$ is a context game, while $\cdot:=[]$ is the terminal object.

Next, $\mathsf{mor}(\mathcal{C})$ and $-\{-\}_{\mathsf{Ty}}$ are defined with $\tot{ Y_j\{[f_k]_{k< j}\}}=\tot{ Y_j}$ and\\
\resizebox{\linewidth}{!}{\parbox{\linewidth}{\vspace{-10pt}
\begin{align*}
&\mathsf{Ctxt}(\DGame_!)([X_i]_{i\leq n},[Y_j]_{j\leq m}) :=\{[f_j]_{j\leq m}\;|\; f_j\in\wstr(\Osat(\Pi_{X_1}\ldots\Pi_{X_n}Y_j\{[f_{k}]_{k<j}\}))\}\\
&Y_j\{[f_k]_{k< j}\}(\sigma_1,\ldots,\sigma_n):=Y_j(\langle \sigma_1,\ldots,\sigma_n\rangle^\dagger;f_1,\ldots,\langle \sigma_1,\ldots,\sigma_n\rangle^\dagger;f_{j-1}).
\end{align*}}}\vspace{-5pt}
Here, $\langle \sigma_1,\ldots,\sigma_n\rangle^\dagger;f_j$ is defined as the usual composition of (winning) strategies on $\tot{ X_1}\& \cdots\&\tot{ X_n}$ and $\tot{ X_1}\Rightarrow\cdots\Rightarrow \tot{ X_n}\Rightarrow \tot{ Y_j}$.

The identities are defined as lists of derelicted copycats. Let us define a strategy $\mathsf{der}_{[X_j]_j,X_i}$ which plays the derelicted copycat on all of $\tot{ X_i}$: $\mathsf{der}_{[X_j]_j,X_i}:=\{s\in P_{\Osat(\Pi_{X_1}\ldots\Pi_{X_{n}}X_i})\;|\; \exists_k s\upharpoonright_{!X_i}\upharpoonright_k\approx_{\tot{ X_i}} s\upharpoonright_{X_i} \}$. We then define $\mathsf{id}_{[X_i]_i}:=[\mathsf{der}_{[X_j]_j,X_i}]_i$ and $\mathbf{p}_{[X_i]_i,[Y_j]_j}:=[\mathsf{der}_{[X_i]_i.[Y_j]_j,X_k}]_k$. Let us define\\ \resizebox{\linewidth}{!}{$\mathsf{Tm}([X_i]_{i\leq n},[Y_j]_{j\leq m}):=\{[f_j]_j\;|\;\textnormal{\begin{diagram}
[X_i]_i& \rTo^{[\mathsf{der}_{[X_i]_i,X_1},\ldots,\mathsf{der}_{[X_i]_i,X_n},f_1,\ldots,f_m]} & [X_i]_i.[Y_j]_j
\end{diagram}}\}.$}\linebreak
Then, we can define $\mathbf{v}_{[X_i]_i,[Y_j]_j}:=[\mathsf{der}_{[X_i]_i.[Y_j]_j,Y_k}]_k$. Note that these are well-defined because of the following claim.
\begin{claim}$\mathsf{der}_{[X_j]_j,X_i}\in\wstr(\Pi_{!X_1}\cdots\Pi_{!X_n}X_i\{[\mathsf{der}_{[X_j]_j,X_k}]_{k\leq i-1}\})$.
\end{claim}
We define  $\langle [f_j]_{j\leq m},[g_k]_{k\leq l}\rangle :=[f_1,\ldots,f_m,g_1,\ldots,g_l]$.
We inductively define the composition of $[X_i]_{i\leq n}\ra{[f_j]_j}[Y_j]_{j\leq m}\ra{[g_k]_k}[Z_k]_{k}$ in $\mathsf{Ctxt}(\DGame_!)$ by
$$[f_j]_j;[g_k]_k:=[\langle f_1,\ldots,f_m\rangle^\dagger ; g_k]_{k},$$

using the usual (co-Kleisli) composition of (winning) strategies on $\tot{ X_1}\Rightarrow \cdots\Rightarrow \tot{ X_n}\Rightarrow (\tot{ Y_1}\&\cdots\&\tot{ Y_m})$ and $\tot{ Y_1}\Rightarrow \cdots\Rightarrow \tot{ Y_m}\Rightarrow \tot{Z_k}$. We note that we can assign to this composition a more precise dependent function type.

\begin{claim}\label{lem:comp}
$[f_j]_j;[g_k]_k$ is a list of winning strategies if $[g_k]_k$ and $[f_j]_j$ are.
\end{claim}
Finally, for $[X_i]_i\ra{[f_j]_j}[Y_j]_j$ and $[g_k]_k\in\mathsf{Tm}([Y_j]_j,[Z_k]_k)$, we can define
$$[g_k]_k\{[f_j]_j\}:=[\langle f_1,\ldots,f_m\rangle^\dagger;g_k]_k.
$$
\begin{remark}Note that, in $\mathsf{Ctxt}(\DGame_!)$, $[A,B]\cong [A\& B]$ if $A$ and $B$ are games (without mutual dependency) and $[]\cong [I]$.
\end{remark}
\vspace{-12pt}
\section{Semantic Type Formers}
\vspace{-7pt}
We show that our CwF supports $1$-, $\Sigma$-, $\Pi$-, and $\mathsf{Id}$-types. We characterise some properties of the $\mathsf{Id}$-types, marking their place in the intensionality spectrum.

\begin{theorem}Our CwF supports $1$-, $\Sigma$- and $\Pi$-type with their $\beta$- and $\eta$-rules.
\end{theorem}
The $1$-type is interpreted by the empty context $[]$ and $\Sigma$-types are just interpreted by concatenation of lists. We define a $\Sigma$-type $\Sigma_{[Y_j]_j}[Z_k]_k\in\mathsf{Ty}([X_i]_{i\leq n})$ as $[Y_j]_j.[Z_k]_k$ for
 $[Z_k]_{k\leq l}\in\mathsf{Ty}([X_i]_{i\leq n}.[Y_j]_{j\leq m})$.

We have already seen $\Pi$-types $\Pi_{[X_i]_{i\leq n}}[Y]:=[\Pi_{!X_1}\cdots\Pi_{!X_n}Y]$ of dependent games. What remains to be defined are $\Pi$-types $\Pi_{[X_i]_i}[Y_j]_j$ of general dependent context games, which can now be reduced to the former, as we have that $\Sigma_{f:\Pi_{x:A}B}\Pi_{x:A}C[f(x)/y]$ satisfies the rules for $\Pi_{x:A}\Sigma_{y:B}C $.
\begin{corollary}Note that this means that $\mathsf{Ctxt}(\DGame_!)$ is in particular a ccc.\end{corollary}

We turn to identity types next, which are essentially defined as those of the domain semantics of \textsf{DTT} \cite{palmgren1990domain}. Interestingly, due to the more intensional nature of game semantics, they acquire a more intensional character, refuting $\mathsf{FunExt}$.

For $[Y_j]_j\in\mathsf{Ty}([X_i]_i)$, define $\mathsf{Id}_{[Y_j]_j}\in\mathsf{Ty}([X_i]_i.[Y_j]_j.[Y_{j'}]_{j'})$ through the intersection of subgames\footnote{Here, we identify a strategy $\sigma$ on $X$ with the subgame $\sigma\cup \{sa\in P_X\;|\; s\in \sigma\}\trianglelefteq X$.} of $\tot{ Y_j}$ for $1\leq j\leq m$:
$$\mathsf{Id}_{[Y_j]_j}([\sigma_i]_i,[\tau_j]_j,[\tau_j']_{j}):=[\mathsf{Id}_{Y_j}]_j([\sigma_i]_i,[\tau_j]_j,[\tau_j']_{j}):=[\tau_j\cap \tau_j']_j.$$
Here, $\tot{ \mathsf{Id}_{Y_j}}:=\tot{ Y_j}$. 
\begin{theorem}
This definition satisfies the $I$-, $E$- and $\beta$-rules for $\mathsf{Id}$-types.
\end{theorem}
For $\mathsf{Id}$-I, $x:A\vdash \mathsf{refl}_t:\mathsf{Id}_B(t,t)$ can be interpreted as the list of strategies $\sem{t}$ but at $\Pi_{\sem{A}}\mathsf{Id}_{\sem{B}}(\sem{t},\sem{t})\trianglelefteq \Pi_{\sem{A}}\sem{B}$, where we write $\sem{-}$ for the interpretation of \textsf{DTT} in our model. We interpret $\mathsf{Id}$-E by sending $H\in \mathsf{Tm}(\sem{A}.\sem{B},\linebreak \sem{C}\{\mathsf{id}_{\langle \sem{A}.\sem{B}}, \mathbf{v}_{\sem{A},\sem{B}}, \mathsf{refl}_{\sem{B}}\rangle\})$ to $J(H)\in \mathsf{Tm}(\sem{A}.\sem{B}.\sem{B}.\sem{\mathsf{Id}_B},\sem{C})$ by playing $H$ between $\sem{A}$, $\sem{\mathsf{Id}_B}$ and $\sem{C}$ rather than $\sem{A}$, $\sem{B}$ and $\sem{C}$.

In addition to being non-extensional (i.e. refuting the principle of equality reflection), these identity types can be said to be intensional in a positive sense.
\begin{theorem}Streicher's Criteria of Intensionality \cite{streicher1993investigations} are satisfied, i.e. \vspace{-4pt}
\begin{enumerate}
\item[(I1)] there exist $\vdash A\;\mathsf{type}$ such that $x,y:A,z:\mathsf{Id}_A(x,y)\not\vdash x\equiv y:A$;
\item[(I2)] there exist $\vdash A\;\mathsf{type}$ and $x:A\vdash B\;\mathsf{type}$ such that  $x,y:A, z:\mathsf{Id}_A(x,y)\not\vdash B\equiv B[y/x]\;\mathsf{type}$;
\item[(I3)] for all $\vdash A\;\mathsf{type}$, $\vdash p:\mathsf{Id}_A(t,s)$ implies $\vdash t\equiv s:A$.
\end{enumerate}
\end{theorem}
(I1) relies on the interpretation of terms carrying intensionality. For instance, we can take $\sem{A}=\B$, where $\mathbb{B}:=\{\ttt,\fff\}$, and evaluate the first and second projections on $\sem{x}=\sem{z}=\bot$ and $\sem{y}=\ttt$. (I2) relies on semantic types having intensional features. We can use $\sem{B}:=(\bot,\fff\mapsto I,\ttt\mapsto\B)$ on the data of (I1). (I3) follows as $[p_i]_i\in\mathsf{Ctxt}(\DGame_!)([],\mathsf{Id}_{[X_i]_i}([f_i]_i,[g_i]_i):=[f_i\cap g_i]_i)$ implies that $p_i=f_i=g_i$ for all $i$, as winning strategies are maximal.

The proofs of (I1) and (I2) also work for the domain model of \textsf{DTT}. (I3) relies on a crucial difference between the domain and games models: winning strategies are maximal, while to account for function types of domains with totality, we cannot assume that total domain elements are maximal. For similar reasons, \textsf{FunExt} is seen to fail in the games model: note that for strict and non-strict constantly $\ttt$ functions $f$ and $g$, we have $[f]\in \mathsf{Tm}([\B],\mathsf{Id}_{[\B]}([f],[g]))$, while $\mathsf{Tm}([],\mathsf{Id}_{\Pi_{[\B]}[\B]}([f],[g])=\emptyset$.
\begin{theorem}\label{thm:id} $\mathsf{FunExt}$ is refuted: for $\vdash f,g:\Pi_{x:A}B$, we do not generally have $z:\Pi_{x:A}\mathsf{Id}_B(f(x),g(x))\vdash\mathsf{FunExt}_{f,g}:\mathsf{Id}_{\Pi_{x:A}B}(f,g). $
\end{theorem}
On the other hand, it turns out that we have the principle of uniqueness of identity proofs \textsf{UIP}, by playing derelicted copycats between $\sem{\mathsf{Id}_A}$ and $\sem{\mathsf{Id}_{\mathsf{Id}_{{A}}}}$.
\begin{theorem}We have $x,y:A,p,q:\mathsf{Id}_A(x,y)\vdash \mathsf{UIP}_A: \mathsf{Id}_{\mathsf{Id}_A(x,y)}(p,q)$.
\end{theorem}

\vspace{-14pt}
\section{Ground Types and Completeness Results}\label{sec:compl}
\vspace{-9pt}
We illustrate how our model of dependent games and winning strategies satisfies a completeness result with respect to the syntax of \textsf{DTT} with $1$-, $\Sigma$-, $\Pi$- and $\mathsf{Id}$-types and finite inductive type families. The precise variant of \textsf{DTT} that these completeness results refer to can be found in the long version of this paper.

We describe a scheme for inductively defining finite type families. Let $A$ be a type not containing any $\Pi$-constructors. Then, we specify a finite inductive definition of a type family $x:A\vdash B\;\mathsf{type}$ by specifying finitely many closed terms $a_1,\ldots, a_n :A$ and distinct symbols $b_{ij}$, $1\leq i\leq n$, $1\leq j\leq m_i$. The idea is that $B$ is a type family, such that $B[a_i/x]$ contains precisely the distinct closed terms $b_{i,1},\ldots,b_{i,m_i}$. These type families are more limited than general inductive definitions as they are freely generated by (finitely many) \emph{closed} terms, while one would allow open terms in the general case. This means that we precisely get the inductive type families that have finitely many non-empty fibres which are all finite types. The prototypical example of such a type family is a calendar in which the type of days depends on the month and year we are in.

We interpret such a definition as specifying $I$- and $E$-rules for $B$:\vspace{4pt}\linebreak
\resizebox{\linewidth}{!}{$\begin{array}{ccc}
\AxiomC{}
\RightLabel{$B$-$I_{i,j}$}
\UnaryInfC{$\vdash b_{i,j}:B[a_i/x]$}
\DisplayProof
& &
\AxiomC{$
x:A,y:B\vdash C\;\mathsf{type}$}
\RightLabel{$B$-$E.$}
\UnaryInfC{$\vdash\mathsf{case}_B: \Pi_{x:A,y:B,z_{11}:C[a_1/x,b_{1,1}/y],\ldots ,z_{nm_n}:C[a_n/x,b_{n,m_n}/y]} C$}
\DisplayProof\vspace{4pt}
\end{array},
$}
together with the $\beta$- and $\eta$-rules, commutative conversions and a rule\footnote{Note that this rule is derivable in presence of a universe.} defining a $\mathsf{exfalso}$ eliminator from $\mathsf{Id}_B(b_{i,j},b_{i',j'})$ for distinct constructors $b_{i,j}$, $b_{i',j'}$ of $B$.

Let $\mathsf{Ctxt}(\DGame_!)_{\mathsf{fin}1\Sigma\Pi\mathsf{Id}}$ be the full subcategory of  $\mathsf{Ctxt}(\DGame_!)$ on the hierarchy generated by $1$-, $\Sigma$-, $\Pi$-, and $\mathsf{Id}$-types and finite dependent games (and substitution), as below. Then we have the following results.
\begin{theorem}[Finite Dependent Games] Finite inductive type families $B$ in context $x:A$, where $B[a_i/x]$ is generated by $\{b_{ij}\;|\;  j\}$, have a sound interpretation in $\mathsf{Ctxt}(\DGame_!)_{\mathsf{fin}1\Sigma\Pi\mathsf{Id}}$: \;\resizebox{0.50\linewidth}{!}{\small
\begin{diagram}
\sem{B}:
\sem{a_i}& \rMapsto & [\widetilde{\{b_{ij}\;|\; j\}_*}], \quad& 
\mathsf{else} & \rMapsto & [\widetilde{\emptyset}_*].
\end{diagram}}
\end{theorem}

\begin{theorem}[$\mathsf{Id}$-free Full and Faithful Completeness]\label{thm:fdef} All morphisms in \\ $\mathsf{Ctxt}(\DGame_!)(\sem{A},\sem{B})$ if $A$ and $B$ do not contain $\mathsf{Id}$-constructors are faithfully definable in \textsf{DTT}.
\end{theorem}
As our interpretation factors faithfully over that of a total finitary PCF, faithfulness follows from (a variation on) the corresponding result for PCF \cite{abramsky2000full}. Definability is proved along the lines of the template of \cite{Abramsky00axiomsfor} and hinges on the decomposition lemma for PCF-games.

Although the completeness properties of the model at the hierarchy with $\mathsf{Id}$-types remain to be studied in detail, we do have the following.

\begin{theorem}[Full and Faithful Completeness for strictly positive $\mathsf{Id}$-types]\label{thm:complid} All morphisms in $\mathsf{Ctxt}(\DGame_!)([],\sem{\Pi_{x:A}\mathsf{Id}_B(f,g)})$ for $x:A\vdash f,g:B$ are faithfully definable in \textsf{DTT}, if $\vdash A\;\mathsf{type}$ and $x:A\vdash B\;\mathsf{type}$ are types built without $\mathsf{Id}$-constructors.
\end{theorem}
This completeness result for $\mathsf{Id}$-types relies on the faithfulness of the interpretation of \textsf{DTT} in our model.
\vspace{-5pt}
\section{Future Work}\vspace{-7pt}
Ultimately, the main goal is a thorough intensional, computational analysis of HoTT \cite{hottbook}. Obvious concrete directions for future work are the following:\vspace{-2pt}
\begin{itemize}
\item breaking $\mathsf{UIP}$, by considering higher dimensional ground types; 
\item examining the phenomena of function extensionality and univalence;
\item study of universes and a more intensional notion of type family;
\item study of (higher) inductive type families and their definability results;
\item establishing completeness results for the full type hierarchy with $\mathsf{Id}$-types;
\item constructing models of \textsf{DTT} with side effects;
\item synthesising strategies from a dependently typed specification;
\item study of a possible embedding of the model in the co-Eilenberg-Moore category $\Gamecat^!$, which might simplify its presentation.
\end{itemize}
\vspace{-3pt}
\subsection*{Acknowledgements}
\vspace{-3pt}
Samson Abramsky was supported by the EPSRC, AFOSR and the John Templeton Foundation.
Radha Jagadeesan acknowledges support from the NSF. Matthijs V\'ak\'ar was supported by the EPSRC and the Clarendon Fund.

\scriptsize
\bibliographystyle{splncs}
\bibliography{tau}

\end{document}